# A technical review of multi-omics data integration methods: from classical statistical to deep generative approaches

Ana R. Baião[1,2], Zhaoxiang Cai[3], Rebecca C Poulos[3], Phillip J. Robinson[3], Roger R Reddel[3], Qing Zhong[3], Susana Vinga[1,2,4#], Emanuel Gonçalves[1,2,#]

1. INESC-ID, 1000-029 Lisboa, Portugal
2. Instituto Superior Técnico (IST), Universidade de Lisboa, 1049-001 Lisboa, Portugal
3. ProCan®, Children's Medical Research Institute, Faculty of Medicine and Health, The University of Sydney, Westmead, NSW, Australia
4. IDMEC, Instituto Superior Técnico, Universidade de Lisboa, Lisboa, Portugal

 # Correspondence: susanavinga@tecnico.ulisboa.pt and emanuel.v.goncalves@tecnico.ulisboa.pt

## Abstract

The rapid advancement of high-throughput sequencing and other assay technologies has resulted in the generation of large and complex multi-omics datasets, offering unprecedented opportunities for advancing precision medicine strategies. However, multi-omics data integration presents significant challenges due to the high dimensionality, heterogeneity, experimental gaps, and frequency of missing values across data types. Computational methods have been developed to address these issues, employing statistical and machine learning approaches to uncover complex biological patterns and provide deeper insights into our understanding of disease mechanisms. Here, we comprehensively review state-of-the-art multi-omics data integration methods with a focus on deep generative models, particularly variational autoencoders (VAEs) that have been widely used for data imputation and augmentation, joint embedding creation, and batch effect correction. We explore the technical aspects of loss functions and regularisation techniques including adversarial training, disentanglement and contrastive learning. Moreover, we discuss recent advancements in foundation models and the integration of emerging data modalities, while describing the current limitations and outlining future directions for enhancing multi-modal methodologies in biomedical research.

# Introduction

Recent advances in high-throughput sequencing technologies have enabled the comprehensive characterization of cellular models across multiple omics layers, encompassing genomics, epigenomics, transcriptomics, proteomics and metabolomics, among others. Multi-omics studies have become commonplace in biomedical research by allowing a holistic representation of biological systems [1,2]. Within the context of precision medicine, multi-omics provides an holistic perspective revealing biological mechanisms at different regulatory layers underlying diseases, enabling the identification of molecular subtypes [3–5], and the discovery of new drug targets [6] and biomarkers for clinical diagnosis, prognosis, and therapeutic response [7–14].

Several consortia have generated invaluable multi-omics datasets and resources, particularly for cancer studies, including TCGA, ICGC [15] and ProCan [16]. Multi-omics data repositories and portals were reviewed in [17–19]. Despite their potential, the integration of these datasets remains challenging due to their high-dimensionality, heterogeneity, and data sparsity [20,21]. Multi-omics datasets often comprise thousands of features, suffering from the "curse of dimensionality" problem, and are generated through diverse sequencing and measurement techniques, leading to inconsistent data distributions across omics [20,22]. Moreover, due to experimental limitations, data quality issues, or incomplete sampling, multi-omics datasets are often unbalanced and incomplete, both at samples and entire modality level [23].

To address these issues, statistical and machine learning models focusing on dimensionality reduction, batch effect correction, and data imputation have been developed **(Figure 1)**. Dimensionality reduction techniques infer joint embeddings that capture the underlying structure and variability across different modalities, facilitating downstream tasks like clustering and classification [24,25]. Batch effect correction mitigates technical biases while preserving critical biological signals [26,27]. Furthermore, imputation techniques enhance data quality through denoising and augmenting datasets to improve robustness and generalizability in downstream analyses [28–30].

Instead of focusing on the types of multi-omics data or biological applications, this review emphasizes the architectural and computational innovations underpinning several methods. Many authors have reviewed multi-omics integration methods, offering diverse perspectives on approaches, challenges, and applications. Several reviews have focused on state-of-the-art statistical and machine learning approaches, categorizing multi-omics integration methods based on different criteria. For instance, some classify these methods by the intrinsic nature of multi-omics experiments, such as vertical, horizontal, diagonal, and mosaic integration *(Figure 1)* [31,32]. Others focus on fusion strategies, including early, intermediate, or late integration [33] and concatenation, model or transformation-based [21]. Other authors emphasized the applications of multi-omics integration methods and the supported omics data types [17,19], with a particular focus on oncology [11,18,34,35]. In Vahabi and Michailidis [36], unsupervised learning methods were reviewed based on their underlying approach, excluding deep learning. Recently, deep learning-based approaches have gained prominence, and many authors reviewed both traditional architectures, emerging trends, their applications, and the omics data types supported [37–46]. More recently, Ballard et al. [46] categorized deep learning approaches based on their architectures and generative capabilities, including variational autoencoders (VAEs) and transformers models.

This review provides a comprehensive technical overview of the methods developed for multi-omics data integration, categorising them into: correlation-based, matrix factorization, probabilistic, network, or kernel-based, and deep learning approaches *(Figure 2)*. Recent advances in the field have shifted the focus from classical statistical approaches to deep learning-based models, particularly generative methods. Therefore, we present a broader understanding of multi-omics integration approaches and their evolution, placing particular emphasis on VAEs, which have gained prominence since 2020 for tasks such as imputation, denoising, and creating joint embeddings of multi-omics data [47–51]. Beyond merely describing VAEs architecture and applications, we explore training strategies and regularisation techniques proposed for adversarial training, cycle consistency, contrastive, and disentangled representation learning. Here, we aim to standardise terminology and provide clarity in a field with a wide range of different methods and definitions. Finally, we discuss promising future directions, including the incorporation of other data modalities and the application of foundation models, which could unlock new possibilities in the multi-modal integration field, enhancing its impact on biomedical research and precision medicine.

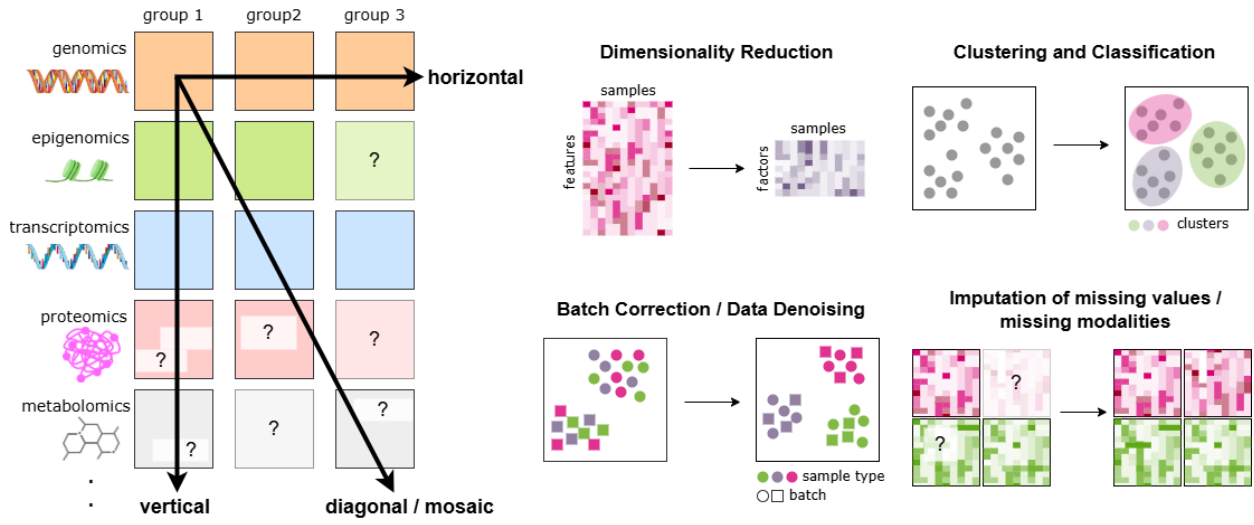

*Figure 1.* Multi-omics data integration. Symbol ? represents missing values or entire missing modalities. <u>Left</u>: Illustration of diverse omics layers (rows) for three different groups of samples (columns), highlighting four integration strategies. Vertical integration combines different omics modalities within the same group of samples; horizontal integration aligns datasets from the same omics layer across different sample groups (e.g., conditions, batches, cellular models), typically addressing batch effect correction; diagonal integration combines distinct omics modalities from different sample groups to explore inter-modality relationships across groups, and mosaic integration leverages overlapping modalities across samples to infer relationships and impute missing modalities. <u>Right</u>: Overview of common tasks in multi-omics analysis. Dimensionality reduction infers low-dimensional embeddings that facilitate downstream tasks like clustering and classification. Batch effect correction ensures that samples cluster based on biological attributes, such as tissue or cell type, rather than technical artifacts, such as sequencing technology or sample source. Imputation addresses missing data, both for randomly missing features and for entire missing modalities, improving the quality and completeness of multi-omics datasets.

## Classical statistical and machine-learning approaches

In this section, we introduce multi-omics integration methods, ranging from correlation and covariance-based techniques to matrix decomposition methods, and probabilistic or bayesian approaches. Lastly, network and kernel-based methods are highlighted. Generally, this considers $M$ different omics matrices $X_i \in \mathbb{R}^{n_i \times p_i}$, $i = 1, \ldots, M$, each with $n_i$ samples and $p_i$ features.

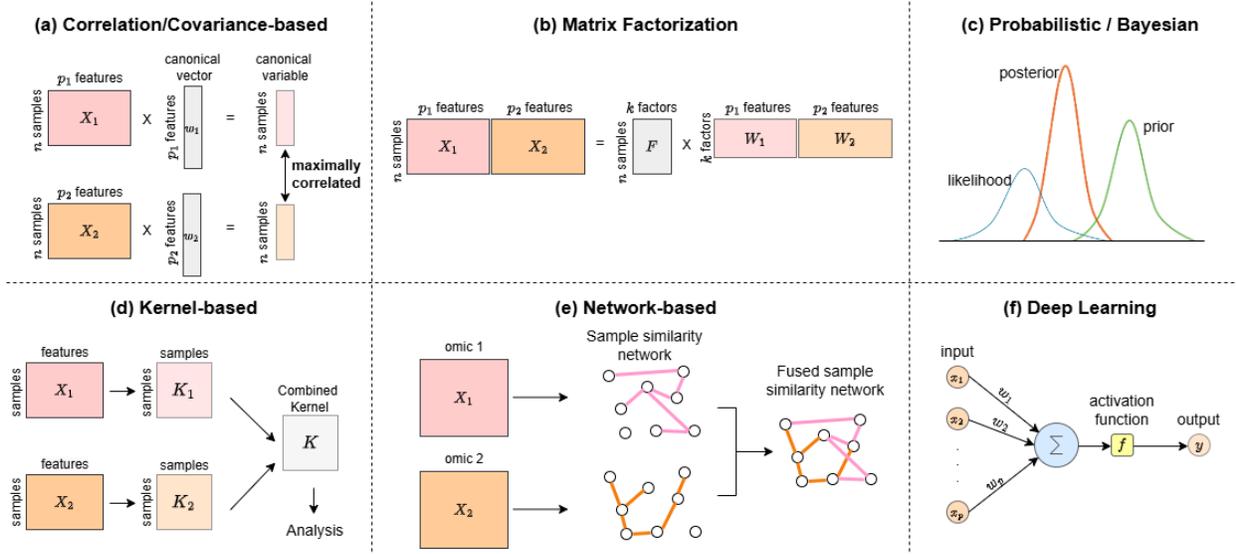

***Figure 2.*** *Schematic representation of the multi-omics integration approaches reviewed. (a) Canonical Correlation Analysis (CCA) framework; (b) Matrix factorization of omics matrices $X_1$ and $X_2$ into a shared matrix $F$ and omics-specific matrices $W_1$ and $W_2$; (c) Bayesian approaches for probabilistic modelling; (d) Multiple Kernel Learning; (e) Similarity Network Fusion (SNF) methodology; (d) Illustration of an artificial neuron, the fundamental unit of neural networks: for each sample vector $x \in \mathbb{R}^p$, each input feature $x_i$ is associated with a weight $w_i$. The neuron computes a weighted sum and the resulting value is passed through an activation function $f$ to produce the output of the neuron $y$.*

## Correlation/Covariance-based methods

**Canonical Correlation Analysis (CCA)** [52] is a classical statistical method designed to explore the relationships between two sets of variables ($M = 2$), with the same set of samples $n$. Two omics datasets, $X_1$ and $X_2$ can be expressed by the following decomposition problem:

$$X_1 = F_1 W_1^T + E_1 \quad \text{and} \quad X_2 = F_2 W_2^T + E_2 \tag{1}$$

where $W_i \in \mathbb{R}^{p_i \times k}$ are omics-specific matrices, $F_i \in \mathbb{R}^{n \times k}$ are omics-specific matrices with $k$ columns of factors or components that explain the co-structure between $X_1$ and $X_2$, and $E_i$ are error terms. CCA aims to find column vectors $w_1 \in \mathbb{R}^{p_1}$ and $w_2 \in \mathbb{R}^{p_2}$ that maximise the correlation between the linear combinations $X_1 w_1$ and $X_2 w_2$:

$$\text{argmax}_{w_1, w_2} \text{corr}(X_1 w_1, X_2 w_2) \tag{2}$$

where $w_1$ and $w_2$ are the first canonical vectors, and $X_1 w_1$ and $X_2 w_2$ are the corresponding canonical variables *(Figure 2a)*.

CCA has proven particularly useful as a joint dimensionality reduction and information extraction method in genomic studies, where multiple types of data, such as DNA copy number or mutation, are often collected from the same set of samples [53–57]. However, a common challenge in multi-omics data analysis is that the number of features typically exceeds the number of observations ($p_i \gg n$), leading to ill-defined problems. Since the optimisation problem in Equation 2 requires the inversion of the covariance matrix, classical CCA cannot be directly applied in such high-dimensional settings. To address this challenge, several CCA extensions were proposed to induce sparsity in the solution of the optimisation problem, including **sparse CCA (sCCA)** [54,55,58], **CCA elastic net (CCA-EN)** [59], or **CCA-sparse group** [56]. To extend the application of CCA-based methods to more than two datasets ($M > 2$), **Regularised Generalised CCA (RGCCA)** and **sparse Generalised CCA (sGCCA)** were proposed [60,61], and are currently one of the most widely used generalisations of CCA to multi-omics data.

**DIABLO** [62] extends sGCCA to a supervised framework. It simultaneously maximises common or correlated information between multiple omics datasets and minimises the prediction error of an outcome. This approach is particularly effective for selecting co-varying modules that explain the response variable, usually phenotypic traits.

Recently, several authors have proposed deep learning-based extensions of traditional CCA to handle nonlinearity and scalability in multi-omics data integration. For example, **SDGCCA** (Supervised Deep Generalised Canonical Correlation Analysis) [63] extends CCA by incorporating neural networks to capture nonlinear cross-data correlations between multiple omics modalities, enabling improved phenotype classification. With the advance of single-cell technologies, **VIPCCA** [64] and **VIMCCA** [65] were proposed for unpaired and paired single-cell data integration, respectively. These methods are based on non-linear canonical correlation analysis, by leveraging deep neural networks and variational inference.

**Partial Least Squares (PLS)** is an alternative approach for data integration that aims to maximize the covariance between components. Several extensions were also proposed to find sparse solutions. For example, **sparse PLS** [66] incorporates a LASSO penalty to enable feature selection, and **sparse Multi-Block Partial Least Squares (sMBPLS)** [67] is applicable to more than two datasets. Many implementations of PLS, which optimise different objective functions with different constraints have been described and reviewed [68].

## Matrix factorization methods

Matrix decomposition is a powerful method for joint dimensionality reduction, condensing datasets into fewer factors to reveal important patterns that can be used to detect disease-associated biomarkers or identifying cancer subtypes, among others.

**JIVE (Joint and Individual Variation Explained)** [69] is considered an extension of Principal Component Analysis (PCA) that decomposes each omics matrix into a low-rank approximation matrix capturing joint variation across omics layers, a low-rank approximation matrix for structured variation specific to each data type, and a residual noise. The joint and individual low-rank approximations are computed by minimising the overall sum of squared residuals. JIVE quantifies the amount of joint variation between data types, reduces the dimensionality of the data, and shows advantages over CCA and PLS by avoiding overfitting.

Non-Negative Matrix Factorization (NMF) is a popular technique to decompose datasets into two non-negative matrices. Several extensions of NMF have been developed to address the specific challenges of multi-omics datasets.

**jNMF** [70] decomposes multiple omics datasets $X_i$ into a shared basis matrix $F \in \mathbb{R}^{n \times k}$ and specific omics coefficient matrices $W_i \in \mathbb{R}^{k \times p_i}$ for each omics **(Figure 2b)**:

$$X_i \approx FW_i \qquad (3)$$

All entries of $F$ and $W_i$ are non-negative. Then, the objective function is formulated as: $\min_{F,W_i} \sum_{i=1}^{M} ||X_i - FW_i||_F^2$, $F, W_i \geq 0$, where $||.||_F$ denotes the Frobenius norm.

**intNMF** [71] is an extension of NMF for clustering of samples using multi-omics data. Once the matrix $F$ have been computed, each sample is associated with one of the $k$ clusters, determined by the highest entry in the matrix.

In the context of single-cell technologies, **LIGER** [72] employs iNMF [73] to be applied to horizontal or diagonal integration problems. LIGER decomposes each omics dataset into dataset-specific weights ($V_i$), shared weights ($W$), and sample specific factors ($F_i$). The objective function is defined as:

$$\min_{F_i, W, V_i} \sum_{i=1}^{M} \left\| X_i - F_i(W + V_i) \right\|_F^2 + \lambda \sum_{i=1}^{M} \left\| F_i V_i \right\|_F^2, F_i, W, V_i \geq 0 \qquad (4)$$

An additional regularisation term is added to the optimisation function to handle omics-specific noise and heterogeneity, allowing the identification of shared cell types across samples and multiple modalities. **UINMF** [74] extends iNMF by adding an unshared weights matrix term to the objective function. This method incorporates features that belong to only one or a subset of the omics datasets, performing mosaic integration.

## Probabilistic-based methods

Matrix factorization is a robust approach for dimensionality reduction but has shown several limitations, particularly in handling missing data and scalability. Probabilistic matrix factorization offers substantial advantages by incorporating uncertainty estimates, allowing for flexible regularisation, and improving biological interpretability through latent structures.

**iCluster** [75] is a joint latent variable model designed to integrate multiple types of omics data, with the purpose of discovering latent cancer subtypes. This method decomposes each omics dataset into a shared latent factor matrix $F \in \mathbb{R}^{n \times k}$ and omics-specific weight matrices $W_i \in \mathbb{R}^{k \times p_i}$:

$$X_i = FW_i + E_i \qquad (5)$$

assuming both the errors $E_i$ and the factor matrix $F$ follow a normal distribution. iCluster derives a likelihood-based formulation of this equation and then applies the expectation-maximisation method to find latent variables $F$ [76]. Clusters are derived by applying the K-means algorithm to matrix $F$. **iClusterPlus** [77] is an extension of iCluster that focuses on modelling different statistical distributions to handle diverse genomic variables but it was criticised for its computational intensity to achieve stable solutions. **iClusterBayes** [78] further extends iClusterPlus by modelling binary genomic variables and RNA sequencing count data using a fully Bayesian inference approach.

**LRAcluster** [79] is another example of clustering probabilistic factor analysis method for continuous and categorical data integration, based on a low-rank probabilistic approach. This method differs from iClusterPlus by using a fast low-rank approximation method to improve the efficiency of parameter estimation to find the latent variables. **moCluster** [80] is also a joint latent model that uses modified consensus PCA [81] for latent variable estimation, offering a stable and efficient alternative to the expectation-maximisation algorithm used in iCluster.

**Multi-Omics Factor Analysis (MOFA)** [82] is an unsupervised multi-omics data integration method that leverages factor analysis within a probabilistic Bayesian framework *(Figure 2b and 2c)*. MOFA also decomposes each omics dataset, as shown in Equation 5, placing prior distributions on all unobserved variables $F$, $W_i$ and $E_i$. This probabilistic approach allows MOFA to accommodate different data distributions and to handle missing values automatically. MOFA employs a two-step regularisation process to handle high-dimensional omics data. First, it identifies which factors are active in each omics data type. Then, it enforces feature-wise sparsity resulting in a small number of features with active weights, enhancing the model's interpretability and its ability to disentangle variation across datasets. By reducing omics data to a low-dimensional factor space, MOFA facilitates various downstream analyses, such as sample classification, clustering, or visualisation.

Unlike other methods such as iCluster, MOFA does not assume normal error distributions and applies different extents of regularisation across factors. Furthermore, MOFA solves the probabilistic Bayesian model by maximising the Evidence Lower Bound (ELBO) [83]. Therefore, a significant advantage in contrast with other methods is its generative capability, allowing it to produce synthetic data. **MOFA+** [84] further enhances this framework, improving its scalability and performance, and broadening its applicability to both bulk and single-cell datasets.

## Kernel-based methods

The previously described models rely predominantly on linear combinations to integrate multi-omics data. Kernel and network-based approaches enable the modelling of nonlinear and complex interactions across diverse biological data layers in a structured way.

Kernel learning approaches [85,86] use kernel functions to map original omics data into higher-dimensional feature spaces. This mapping is represented by a kernel matrix $K$ that represents the similarities between all pairs of data points, computed as the inner product between their representations in the feature space $K_{i,j} = k(x_i, x_j) = <\phi(x_i), \phi(x_j)>$, where $\phi$ maps the original data to the feature space. The kernel function $k(x_i, x_j)$ is the only required definition for the kernel method. In multi-omics integration problems, a kernel matrix is computed for each omics layer. Then, multiple kernel learning (MKL) combines them to produce an integrated final kernel matrix by minimising an objective function. The final kernel matrix is used for pattern analysis and clustering *(Figure 2d)*. With kernel learning methods, the problem of data integration is transformed into kernel integration in the sample space rather than the heterogeneous feature space. As a result, the optimisation problems are independent of the number of features, and these methods are called dimension-free.

Several methods have been proposed for multi-omics integration using this type of approach, for example, **rMKL-LPP** [87] uses a linear combination of kernels constructed using an objective function based on the Locality Preserving criterion and **web-rMKL** [88] is the intuitive interface to run the model on a web server. **pairwiseMKL** [89] is a time and memory-efficient version of MKL with applications to drug response prediction.

## Network-based methods

Network-based methods leverage graphs and more advanced network structures to represent omics data and their relationship. By modelling omics data as a network, these methods capture topological structures and interactions, offering a way to infer functional modules, identify key biomarkers, and reveal hidden biological pathways. **Similarity Network Fusion (SNF)** [90] is a popular network-based method for multi-omics integration. SNF constructs patient similarity networks for each omics layer, where nodes represent samples and edges represent their similarity, and iteratively fuses them into a single network using a nonlinear combination method based on message passing theory *(Figure 2e)*. This approach results in a unified similarity network that captures shared patterns across omics layers, enhancing its utility in subtype identification for a broad range of diseases [91–94]. These similarity-based methods have improved runtime since they mainly depend on the number of samples rather than the number of features. However, SNF does not distinguish between data types, and relies on Euclidean distance to calculate sample similarity, which may not fully capture complex relationships in high-dimensional omics data. To address these limitations, several extensions of SNF, along with deep learning-based methods, have been developed recently [95–99].

**NEMO** [100] is a popular similarity network-based method that handles unmatched samples without needing data imputation. For each omics dataset a patient similarity matrix is built using a radial basis function kernel and converted by adjustments based on local neighbourhoods. Then, an average relative similarity matrix that captures the information across all omics layers is obtained. Spectral clustering is used to reveal subtypes based on this matrix. NEMO was able to identify patient subgroups that showed significant differences in terms of survival, and achieved superior performance compared to other nine clustering methods.

Network-based models leverage known interaction networks and molecular pathways, providing biological interpretable frameworks for multi-omics integration. However, their adaptability is limited by their reliance on predefined networks.

# Deep learning approaches

Deep learning approaches have emerged as powerful tools for the multi-omics integration field, offering greater flexibility to integrate high-dimensional and diverse data types and to learn nonlinear, complex patterns from data. In the following section, we explore various neural network architectures for multi-omics integration, ranging from non-generative to generative approaches, placing particular focus on VAEs.

## Non-generative models

Non-generative approaches focus on learning direct mappings or relationships between input features and outputs, often prioritising tasks like classification, regression, or dimensionality reduction.

### Feed Forward Neural Networks

The feed-forward neural network (FFNN) is the most common neural network architecture, consisting of fully interconnected layers of neurons. The individual neurons compute a weighted sum of their input, followed by the application of an activation function and propagating it forward to the next layer *(Figure 2f)*. The activation function typically introduces non-linearity, increasing the expressive power of the network yielding more complex relationships. These models are trained to minimise a loss function using optimisation techniques, such as backpropagation. Despite being computationally intensive and having a high number of parameters, neural networks have demonstrated significant potential in capturing and modelling non-linear relationships across diverse omics datasets [101,102].

**MOLI** [103] is an example of a supervised FFNN designed for drug response prediction. This method employs separate subnetworks for each omics layer, effectively serving as feature extractors. The outputs from these subnetworks are concatenated into a unified representation, which is then passed to a final neural network for classification. MOLI incorporates a binary cross-entropy loss and a triplet loss function for model training. It has been tested on several datasets for classifying patients as responders or non-responders to specific cancer drugs, achieving notable results compared to other concatenation methods. This architecture has also been applied to other tasks, such as synergistic drug combination prediction [104,105], survival analysis [106] or trajectory inference [107].

## Convolutional Neural Networks

Convolutional neural networks (CNNs) are an architecture particularly successful for image and audio data due to their ability to learn spatial hierarchies and capture local patterns through convolutional operations. Although CNNs have some applications in multi-omics data integration [108–111], their use is relatively limited, probably due to the lack of an inherent structural organisation in omics tabular data, unlike the clear grid-like structure present in image data, and thus they will not be detailed in this review.

## Graph Neural Networks

Graph neural networks (GNNs) are a powerful framework for processing data structured as graphs, making them particularly valuable in biological research where entities are intrinsically linked, such as in protein-protein interactions (PPIs) or gene regulatory networks. Graph Convolutional Networks (GCNs) are the most dominant GNNs and introduce convolution operations to the graph structure. **MOGONET** [112] is designed for supervised multi-omics data integration and classification by constructing a sample similarity network for each omics type and leveraging GCNs to predict labels based on individual modalities. **scMoGNN** [113] is an example of a framework that leverages GNNs for single-cell multi-omics data integration to tackle modality prediction, matching, and joint embedding tasks.

While methods like MOGONET focus on sample similarity networks, they do not incorporate biological interaction data, such as PPIs, which could provide additional meaningful context. For instance, Zhuang et al. [114] proposed a GCN method for disease classification integrating transcriptomics and proteomics data with PPI networks.

## Autoencoders

Autoencoders are an unsupervised deep learning model, widely used for dimensionality reduction tasks and feature extraction. This model architecture leverages neural networks to compress the input data into a lower dimensional latent space via an encoder and attempts to reconstruct it back to the original space through a decoder. Several autoencoder-based models have been developed for multi-omics integration for cross-modality translation tasks [115,116] or for joint dimensionality reduction, where the latent space is used for various downstream tasks like disease prognosis and subtyping [117–122], clustering [123,124], synergistic drug combination prediction [125], or batch correction [126,127].

Among several extensions of autoencoders, VAEs are the most prominent for multi-omics data analysis due to their probabilistic framework and generative capability, which will be detailed in the next section.

## Variational Autoencoders

In contrast with non-generative approaches, generative models aim to learn the underlying data distribution itself, enabling the generation of new data points by sampling from the learned distribution. Deep generative models (DGMs) have revolutionised the field of molecular biology, including multi-omics data integration [128]. The majority of DGMs published and reviewed here are based on VAEs [129]. However, other frameworks including generative adversarial networks (GANs) [130] and, more recently, generative pretrained transformer (GPT) approaches [131] have been proposed.

In a common Bayesian approach, it is assumed that each sample vector $x \in \mathbb{R}^p$, with $p$ features, is generated by a latent vector $z \in \mathbb{R}^k$, where $k \ll p$. Each latent vector $z$ is drawn from a prior distribution $p_\theta(z)$ and the new observation is generated from the conditional likelihood distribution $p_\theta(x|z)$, where $\theta$ are the parameters of the generative network (decoder). Therefore, the marginal likelihood is estimated as $p_\theta(x) = \int p_\theta(x,z)\, dz = \int p_\theta(x|z)\, p_\theta(z)\, dz$. This is a computationally intractable problem and other approximation inference methods need to be employed to efficiently estimate the model parameters.

VAE [129] provides a principled way for performing variational inference that can be divided into two main processes. The primary goal is to approximate the true posterior distribution $p_\theta(z|x)$, which maps the input data $x$ in the low-dimensional latent space $z$, using a variational posterior $q_\phi(z|x)$, where $\phi$ are the parameters of the encoder (inference network). Therefore, to infer the data likelihood, we can take the expectation with respect to $q_\phi(z|x)$ and decompose as:

$$\log p_\theta(x) = \mathbb{E}_{q_\phi(z|x)}[\log p_\theta(x)] = \mathbb{E}_{q_\phi(z|x)}\left[\log \frac{p_\theta(x,z)}{q_\phi(z|x)}\right] + D_{KL}\left(q_\phi(z|x)||p_\theta(z|x)\right) \quad (6)$$

Where $D_{KL}$ is the Kullback-Leibler divergence, which quantifies how well the variational distribution $q_\phi(z|x)$ approximates the true posterior and is always positive. Consequently, the first term of the equation is named ELBO, a lower bound on data likelihood. The parameters $\phi$ and $\theta$ of the encoder and decoder distributions, respectively, can then be optimised by maximising the ELBO:

$$\log p_\theta(x) \geq ELBO_{VAE}(x,\phi,\theta) = \mathbb{E}_{q_\phi(z|x)}[\log p_\theta(x|z)] - D_{KL}\left(q_\phi(z|x)||p_\theta(z)\right) \quad (7)$$

The ELBO consists of two terms: the reconstruction term, which measures how accurately the decoder can reconstruct the input data from the latent representation $z$, and the KL divergence term, which regularises the model by minimising the divergence of the learned variational posterior from the prior $p_\theta(z)$. The variational posterior $q_\phi(z|x)$ typically follows a standard gaussian distribution. However, in single-cell data applications, other probability distributions, such as negative binomial, can be considered to handle the sparsity and count nature of the data [132,133].

The overall loss function of a VAE is the negative ELBO, which aims to minimise the reconstruction error, typically using the mean squared error (MSE), and the KL divergence term:

$$L_{VAE} = L_{reconstruction} + \lambda L_{KL} \quad (8)$$

where $\lambda$ is a hyperparameter to weight the KL divergence term and balance between data reconstruction and model regularisation.

To train a VAE, the encoder network processes input data and outputs two layers representing the mean μ and the standard deviation σ of the variational posterior $q_\phi(z|x)$. The reparameterization trick is used to make sampling differentiable, allowing backpropagation, by generating $z$ as $z = \mu + \sigma\varepsilon$, where ε is drawn from a standard normal distribution. Finally, the latent variable $z$ is the input of the decoder network that will reconstruct the input data *(Figure 3a)*.

VAEs are widely used for multi-omics data analysis, particularly for single-cell experiments with high-dimensional and incomplete data, due to their flexible designs and ability to balance dimensionality reduction and generative capabilities. Over the past years, numerous methodological improvements have been proposed to enhance VAEs performance in multi-omics integration tasks. In the following sections, strategies used to optimise VAEs and their functionality for multi-omics data integration and single-cell applications are detailed. Table 1 provides an overview of the different generative methods reviewed in this paper, highlighting the applied strategies and the types of omics data integrated by the studies proposing each model.

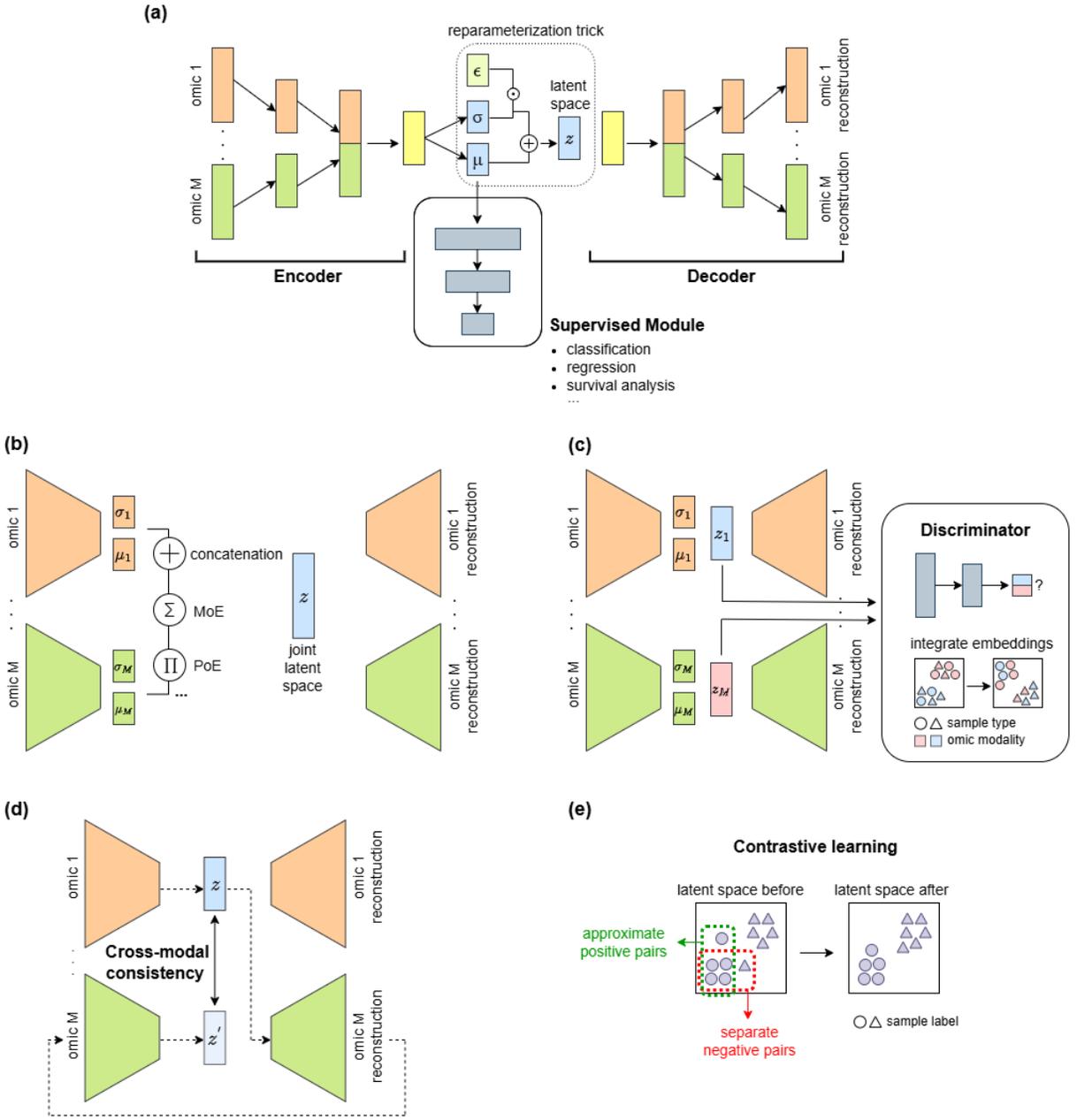

***Figure 3.*** *VAEs architectures. (a) VAE with a supervised module for task-specific supervision. Each rectangle represents a fully connected block. Data from two omics are concatenated in the second hidden layer. The parameters $\mu$ and $\sigma$ represent the mean and standard deviation of the learned posterior distribution, and $\epsilon \sim \mathcal{N}(0,1)$. The reparameterization trick is depicted in the dashed box. (b) Multimodal VAE architecture, highlighting three strategies to build the joint latent space: concatenation, mixture-of-experts (MoE), and product-of-experts (PoE). (c) Adversarial Training strategies in VAEs to align the latent spaces of different omics modalities. (d) Cross-modal cycle consistency. (e) Contrastive learning to self-supervise VAEs by gathering positive pairs and separating negative pairs.*

**Table 1.** *Overview of deep generative models for multi-omics data integration. CyTOF: cytometry by time of flight; CNV: copy number variation; scATAC-seq: single cell sequencing assay for transposase-accessible chromatin; scRNA-seq: single cell RNA-sequencing; snRNA-seq: single nucleus RNA-sequencing; snmC-seq: single nucleus methylation sequencing.*

| Name | Method | Omics modalities |
|---|---|---|
| **OmiVAE / XOmiVAE** [134,135] | VAE with a supervised module for classification / extension incorporating Deep SHAP for biological interpretability | gene expression, DNA methylation |
| **scMVAE** [136] | MVAE with three strategies of joint learning (direct concatenation, neural network, PoE) | scRNA-seq, scATAC-seq |
| **SCIM** [47] | VAE with adversarial training to distinguish between omics technologies based on the latent space | scRNA-seq, CyTOF |
| **MMD-VAE** [48] | VAE with MMD instead of KL regulariser and a classifier network for supervised learning | DNA methylation, CNV, mRNA or RNAseq |
| **OmiEmbed** [137] | VAE with a supervised module for classification, regression, and survival prediction | miRNA, gene expression, DNA methylation |
| **scMM** [132] | MVAE with MoE and a pseudo-cell generation strategy for model interpretability | scRNA-seq, scATAC-seq or surface protein |
| **DCCA** [138] | Separate VAEs mutually supervised by cross-omics cycle attention | scRNA-seq, scATAC-seq |
| **Cobolt** [139] | MVAE with PoE for the integration of data from multi and single-modality platforms | scRNA-seq, scATAC-seq |
| **omicsGAN** [140] | GAN to integrate two omics modalities and their interaction networks | mRNA, miRNA expression |
| **DAVAE** [141] | VAE with a domain-adversarial regulariser to distinguish between original batch label based on the latent space | scRNA-seq, snRNA-seq or scATAC-seq or spatial transcriptomics |
| **scMVP** [142] | MVAE with GMM prior, attention-based channels, and two intra-modal consistency modules to align each reconstructed/imputed omics | scRNA-seq, scATAC-seq |
| **Multigrate** [49] | MVAE with an extra shared decoder, PoE strategy, and an additional MMD regulariser to minimise the distance between joint latent representations for pairs of datasets | scRNA-seq, scATAC-seq or surface protein |
| **GLUE** [143] | Omics-specific VAEs and a graph VAE guided by a prior knowledge-based graph; adversarial training to distinguish between omics based on cell embeddings | scRNA-seq, scATAC-seq, snmC-seq |
| **Portal** [144] | Encoder and GAN: modality-specific encoders, cross-modal generators, and discriminators; additional | scRNA-seq, scATAC-seq or snRNA-seq |

| | regularisers for cross-modal embeddings and samples consistency, and intra-modality reconstructions | |
|---|---|---|
| **sciCAN** [145] | Encoder and GAN: one discriminator to distinguish between omics latent spaces, one discriminator to distinguish between real and cross-modal generated data, cross-modal embeddings cycle-consistency | scRNA-seq, scATAC-seq |
| **scVAEIT** [146] | VAE with conditional variational inference using missing masks to integrate and impute multimodal datasets with mosaic measurements | scRNA-seq, scATAC-seq, surface proteins |
| **Matilda** [147] | VAE with a supervised module for classification of single-cell data | scRNA-seq, scATAC-seq, surface protein |
| **JAMIE** [50] | VAEs using cross-modal correspondence and correlation-based latent aggregation to build aggregate latent spaces; SHAP values for feature importance analysis | scRNA-seq, scATAC-seq |
| **MultiVI** [148] | MVAE using distributional average and penalization to mix latent embeddings; symmetric Jeffrey's divergence term to minimise the distance between latent embeddings of each modality; adversarial training to distinguish between batches or modalities based on the shared latent space | scRNA-seq, scATAC-seq, surface protein |
| **scDisInFact** [149] | VAE with an additional MMD term to disentangle condition-associated signals from batch effects | scRNA-seq |
| **MIDAS** [51] | MVAE with PoE that employs self-supervised modality alignment, information-theoretic latent disentanglement, and masking techniques to handle missing modalities | scRNA-seq, scATAC-seq, surface protein |
| **scCross** [150] | Omics-specific VAEs with a FFNN aligner to build a joint latent embedding; one discriminator to distinguish between omics based on the shared latent space and omics specific discriminators to distinguish between original and reconstructed data to ensure consistency | scRNA-seq, scATAC-seq, snmC-seq |
| **MOSA** [151] | Conditional MVAE with direct concatenation of latent vectors and a self-supervised contrastive loss | CNV, gene expression, proteomics, methylomics, metabolomics, drug response, CRISPR-Cas9 |

## Maximum Mean Discrepancy Regulariser

Despite the success of variational autoencoders, the usual ELBO-based loss function can fail to learn (1) an inference distribution $q_\phi(z|x)$ that approximates the true posterior $p_\theta(z|x)$, and (2) meaningful or informative latent features [48,152]. Inference failures can occur due to an imbalance in optimisation of the ELBO or modelling bias. When the input data is high-dimensional compared to the latent space, modelling errors can be amplified. This causes the model to prioritise data reconstruction over learning a distribution that approximates the true posterior, potentially leading to poor generalisation and overfitting. Additionally, VAEs can reconstruct input data without relying on the latent variables. This results in the model ignoring these variables, making them uninformative and failing to capture meaningful information about the input data.

To address these issues, several models [48,49,148,153] have employed the Maximum Mean Discrepancy (MMD) [154] in the loss function as a regulariser, instead of the KL divergence. The MMD quantifies the distance between two distributions $p(z)$ and $q(z)$ and can be defined as:

$$MMD(p(z)||q(z)) = \mathbb{E}_{p(z),p(z')}[k(z,z')] + \mathbb{E}_{q(z),q(z')}[k(z,z')] - 2\,\mathbb{E}_{q(z),p(z')}[k(z,z')] \quad (9)$$

Where $k(z,z')$ can be any positive definite kernel. A popular choice is the Gaussian kernel $k(z,z') = e^{-\frac{||z-z'||^2}{2\sigma^2}}$. MMD-based regulariser estimate the divergence by how different the moments of the two distributions are, where $MMD(p(z)||q(z)) = 0$ only if $p(z) = q(z)$. MMD-VAEs can capture complex relationships in the data with flexible kernel choices, improving model adaptability and overall performance in integrating diverse omics data types.

Several models reviewed in the following sections incorporate MMD terms into their loss functions. For instance, **MMD-VAE** [48] designed for ovarian cancer analysis replaces the KL divergence with an MMD regulariser. **Multigrate** [49] and **MultiVI** [148] add MMD terms to the standard VAE loss to ensure alignment and consistency across omics modalities. On the other hand, **scDisInFact** [149] introduces a MMD term to ensure disentanglement of latent factors, improving interpretability.

## Supervised learning tasks with VAEs

In a standard unsupervised VAE, the bottleneck layer is designed to extract the most essential features for accurate input data reconstruction. However, these extracted features are often general and may not be relevant to particular downstream analysis. In multi-omics and cancer research, several approaches have extended VAE architectures to incorporate supervised modules for downstream tasks, such as classification, regression, or prognosis prediction.

In VAE-based models with supervised modules, each omics layer is processed with specific encoders to generate feature vectors. These vectors are concatenated and encoded into a unified multi-omics vector, which provides the mean and variance of the inferred distribution. The output vector $\mu$ is then connected to a neural network to perform a specific downstream task **(Figure 3a)**. This additional network introduces a task-specific regulariser, by summing the loss of the downstream task (e.g., mean squared error or binary cross-entropy) into the overall VAE loss function. This combined loss ensures that the latent factors extracted by the VAE not only accurately reconstruct the input data but are also informative for the specific supervised tasks.

One such model is **OmiVAE** [134], which integrates gene expression and DNA methylation data to classify pan-cancer tumour samples. **XOmiVAE** [135] extends this model by incorporating explainability into the VAE framework. This method uses Deep SHAP [155] to provide the contribution of each individual feature and omics latent dimension for the cancer classification task. Therefore, XOmiVAE adds a layer of biological interpretability, which is particularly useful for evaluating how different molecular features contribute to disease classification and for the identification of potential biomarkers.

**OmiEmbed** [137] extends the previous models by introducing a multi-task learning with different downstream tasks including cancer classification, demographic and clinical features reconstruction, and survival prediction. In this case, the overall loss function of the downstream modules is the weighted sum of all downstream losses. OmiEmbed introduces a multi-task training strategy where the low-dimensional embedding is shared across various downstream supervised tasks, and where the information from different tasks is leveraged to improve the overall model performance across all tasks.

**MMD-VAE** [48] proposes a novel integrated multi-omics analysis of ovarian cancer using a VAE that supports tri-omics data analysis, including molecular subtypes clustering, classification and survival analysis. This model uses the MMD regulariser instead of the KL divergence to address the issues mentioned in the previous section. The results show that MMD-VAE outperforms VAE in most omics datasets.

In the context of single-cell data analysis, **Matilda** [147] is a unified VAE framework that integrates multiple data modalities and performs multiple tasks, such as cell type classification and feature selection.

VAE-based models with supervised modules significantly advance omics analysis by integrating generative modelling with supervised learning. This approach enhances the extraction of biologically relevant features and optimises classification or regression performance, addressing limitations of traditional unsupervised methods.

## Inferring joint latent representations using MVAEs

Multimodal VAEs (MVAEs) are a common approach for multi-omics data integration, where each omics modality is assigned its own encoder-decoder, and a shared latent space is constructed. Various strategies exist for combining the latent variables from each modality's encoder into a unified latent representation. These strategies include, among others, direct concatenation or probabilistic methods such as the mixture of experts (MoE), product of experts (PoE) [156], and, more recently, mixture-of-product-of-experts (MoPoE) [157] **(Figure 3b)**.

In MoE approaches, the joint variational posteriors for the $M$ individual modalities is defined as $q_\phi(z|x_{1:M}) = \sum_{i=1}^{M} \alpha_i q_{\phi_i}(z|x_i)$, usually with $\alpha_i = \frac{1}{M}$. The resulting ELBO is the weighted average of the ELBO individual modalities:

$$ELBO_{VAE-MoE} = \frac{1}{M} \sum_{i=1}^{M} \left\{ \mathbb{E}_{z_i \sim q_{\phi_i}(z|x_i)} [\log p_\theta(x_{1:M}|z_i)] - D_{KL}\left(q_{\phi_i}(z|x_i)||p_\theta(z)\right) \right\} \quad (10)$$

Where $x_i$ and $z_i$ are sample and latent vectors for modality $i$, respectively.

The PoE is an alternative approach that infers the joint posterior as the product of the variational posteriors of the individual modalities $q_\phi(z|x_{1:M}) = \prod_{i=1}^{M} q_{\phi_i}(z|x_i)$.

**scMVAE** [136] is a vertical integration method composed of a multimodal encoder and single-modal encoders/decoders for each omics modality. The multimodal encoder employs three strategies for constructing a joint representation: direct concatenation of the input data, a neural network to combine the features extracted by a sub-encoder network for each omics layer, and PoE. Single-modal encoders handle tasks such as data normalisation, denoising, and imputation, ensuring that each individual modality's input is properly processed before being combined. This model includes a Gaussian Mixture Model (GMM) as the prior to generate highly realistic samples by learning more disentangled and interpretable latent representations.

**scMM** [132] extends scMVAE, allowing for cross-modal generation, employing an independent encoder network for each modality and a MoE strategy to build the joint latent representation. This model enhances interpretability by sequentially generating pseudo cells from different latent values in one dimension with remaining fixed values and calculating the Spearman correlation for each latent dimension and set of features in each modality. This approach allows the identification of features that are strongly associated with each latent dimension, improving the interpretation of the results. Additionally, scMM infers latent variables that can reconstruct the probability distributions not only for their own modalities but also for others, having the capability to generate missing data from one modality using data from another omics in both directions.

**Cobolt** [139] is a MVAE with a PoE approach designed for the integration of single-cell data from joint and single-modality platforms. This is particularly important because single-modality data is much more prevalent than joint-modality data in terms of both quality and quantity. Cobolt addresses this by learning a unified latent representation of the cells, regardless of whether the cell data comes from single or joint modalities, making it a versatile tool for batch correction and clustering.

**Multigrate** [49] also employs a PoE approach to effectively combine the posteriors of different modalities to perform mosaic integration and is trained conditionally on a set of study labels (samples, experiments across labs or sequencing technologies). One key innovation is the use of an additional MMD loss term using multi-scale radial basis kernels, which minimises the distance between joint latent representations for pairs of datasets. The model architecture features modality-specific encoders and decoders, along with a shared decoder that processes the joint latent representation. This dual-decoder design ensures both common biological patterns and modality-specific variations are captured. Multigrate allows multi-modal reference building, mapping of new query data into a reference atlas using transfer learning [158], and imputation of missing modalities.

**MultiVI** [148] is a MVAE from the scvi-tools library [159] built on earlier VAE-based methods, like scVI [160] and totalVI [133] and conceptually similar to Cobolt. However, MultiVI is trained conditionally on a set of covariates, uses tailored noise models for each omics modality and combines the information from different modalities into a shared latent representation using a distributional average and penalization strategy. For example, considering two omics modalities sample vectors $x_1$ and $x_2$, the shared latent space is defined as:

$$z = w_1 z_1 + w_2 z_2 \quad \text{with} \quad w_1 + w_2 = 1 \tag{11}$$

Where $w_1$ and $w_2$ are the weights for each modality, sample-specific or the same for all samples, also optimised during training. For samples with only one available modality, the latent space is drawn directly from the representation for which data are available. To obtain a latent space that reflects both modalities, an additional term is added to the common VAE loss to minimise the distance between the two latent representations using symmetric Jeffrey's divergence:

$$L_{symmKL} = D_{KL}(q(z_1)||q(z_2)) + D_{KL}(q(z_2)||q(z_1)) \tag{12}$$

An alternative penalization scheme replacing the symmetric KL divergence by an MMD penalty was also explored by the authors.

The projection of multi-omics data into a common latent space using MVAEs has become the most prevalent strategy for integration. The unified latent representation facilitates a range of downstream analyses, including clustering and visualization, by capturing shared biological patterns across omics layers. However, such embeddings inevitably attenuate omic-specific patterns, potentially hiding unique insights from individual omics datasets. To address this limitation, several models have been designed with cross-learning approaches, which explicitly retain omics-specific patterns while leveraging shared information.

## Cross-learning approaches

The authors of scMVAE have extended this model to address some limitations, such as the projection of the multi-omics data into a common embedding space and the restriction that all modalities need to be present during training and inference. Therefore, **DCCA** [138] was proposed to combine multiple modalities into a separate but coordinate embedding space that preserves the unique characteristics of each modality. DCCA processes each omics modality with a separate VAE that can learn from each other with mutual supervision through cross-omics attention transfer. In DCCA, a well-trained network on one modality acts as a teacher to guide the training of a student network on another modality. The model employs an additional term to the loss of each VAE to minimise the differences between the latent variables of each VAE, ensuring that the embeddings from both networks are aligned:

$$L_{DCCA-alignment} = \beta \sum_{i=1}^{k} ||z_2{}^i - z_1{}^i||_2 \tag{13}$$

Where $z_1$ and $z_2$ are $k$-dimensional latent vectors for two different omics modalities, and $\beta$ the weight of the added term. The results demonstrate that scATAC-seq data generated from scRNA-seq data achieved correlations of 0.9 or higher with the true scATAC-seq data across two independent datasets.

**JAMIE** [50] is a VAE framework for di-omics integration and imputation. This model incorporates an optional cross-modal correspondence matrix $F$ to handle partially aligned samples. JAMIE employs encoders to transform each modality into separate latent spaces, which are then aggregated using correlation-based latent aggregation using matrix $F$. This model is optimised via a combination of loss functions: the common KL divergence and reconstruction terms, and combination and alignment terms. The combination term enforces similarity between the aggregate matrices and the separate latent spaces. The alignment loss shapes the aggregated latent spaces to enforce the similarity of cross-modal correspondent cell representations. JAMIE's architecture supports imputation from one modality to another, phenotype prediction, and the prioritisation of input features for cross-modal imputation using SHAP. Its ability to adaptively learn correspondences and generate a reusable latent space makes JAMIE a versatile tool for multi-omics integration and analysis.

## Adversarial training strategies for VAEs

With the advance of DGMs, GANs have emerged as an innovative approach for multi-omics data integration, leveraging their unique adversarial training strategy. Their generative capability allows the production of realistic data and the discriminative ability to differentiate synthetic data from real data, enabling robust modelling of complex data distributions.

GANs [130] architecture consists of two competing neural networks jointly optimised: a generator network $G$ that learns how to transform the input noise distribution $p(z)$ into the observed data distribution $p(x)$, and the discriminator network $\mathcal{D}$ that learns to distinguish between the real data $x$ and the synthetic data generated $G(z)$. Therefore, the GAN training process can be formulated as a zero-sum minimax game with the following objective function:

$$\min_{G} \max_{\mathcal{D}} \mathbb{E}_{x \sim p(x)}[\log \mathcal{D}(x)] + \mathbb{E}_{z \sim p(z)}\left[\log\left(1 - \mathcal{D}\big(G(z)\big)\right)\right] \tag{14}$$

Where $\mathcal{D}(x)$ is the discriminator's output, representing the probability that the input data $x$ comes from the real data distribution, and $\mathcal{D}(G(z))$ is the probability that the probability that the generated data $G(z)$ resembles real data. Consequently, $\mathcal{D}$ tries to make $\mathcal{D}(G(z))$ near 0 and $G$ tries to make $\mathcal{D}(G(z))$ near 1.

**omicsGAN** [161] leverages Wasserstein GANs to integrate two omics modalities and their interaction networks to learn inter-modality relationships. The generator is trained and updated to synthesise data for one omics modality using data from the other modality and the adjacency matrix of the interaction network. The discriminator is adversarially trained to distinguish between real and synthetic data. The results showed that the integrated data generated by the model had better performance in cancer outcome classification and survival prediction compared to the original data on breast, lung, and ovarian TCGA cancer datasets.

While GANs have flexible modelling and enhanced data distribution learning, they also come with challenges. These include the complexity of training dual networks, scalability issues when handling a larger number of omics modalities, or the need for large sample sizes to achieve stable training and meaningful results. To address these limitations, the most common approach in multi-omics analysis is to integrate adversarial training approaches into VAE frameworks as additional components to regularise the latent space or the decoder reconstruction.

In these VAE–based models, the discriminator is commonly employed to distinguish between omics modalities or technologies based on samples from the latent space *(Figure 3c)* or on reconstructed samples from cross-modal decoders. The adversarial penalty encourages the model to better align different modalities in the latent space or to learn decoders that allow for accurate cross-modal predictions well aligned with the intra-modal predictions.

Considering a discriminator acting on samples from the latent sample, its goal is to maximise the probability of correctly identifying the original omics modality or batch a sample comes from, while the encoder and the decoder are trained to fool the discriminator by producing samples that are indistinguishable. In a successful integration, the latent space of each modality should be integrated well such that they are indistinguishable from each other, with corresponding cells across all modalities represented in close proximity. During training, this can be achieved by incorporating an adversarial penalty term into the standard VAE loss function.

**SCIM** [47] uses adversarial training to match cells from a source technology to cells in one or multiple target technologies in two main steps, allowing for diagonal integration tasks. It uses separate encoders for each technology to build a technology-invariant latent space. Through adversarial training, the encoders are encouraged to produce similar latent representations, effectively merging the separate latent spaces into one integrated space. This method minimises reconstruction loss, ensuring that important biological information is preserved while making data from different technologies comparable. The final output is a cohesive latent space that facilitates reliable cross-technology analysis. Then, cells are paired across different technologies via their latent representations using a version of the fast bipartite matching algorithm.

**DAVAE** [141] leverages domain-adversarial and variational approximation techniques. The model employs a shared encoder, requiring a common set of input features, and adversarial training to address batch effects. By introducing a domain classifier that predicts the batch or modality of samples from the latent representation, the model is trained to fool the classifier, ensuring that batch effects are minimised in the final integrated representation.

**GLUE** [143] is designed for multi-omics diagonal integration tasks through the use of graph-guided embeddings. GLUE uses omics-specific VAEs to learn sample latent spaces with the same dimension for each omics modality. A graph VAE is also used to learn feature latent spaces for each omics using guidance graphs that captures prior knowledge about regulatory interactions across modalities: vertices represent features from different omics layers, while edges represent known regulatory relationships. The feature and sample latent spaces created are combined via inner product to reconstruct omics data. Furthermore, a discriminator is employed to align the cell embeddings across different modalities via adversarial training using the multiclass classification cross entropy, ensuring proper alignment and batch correction. GLUE is particularly powerful for tasks such as dimensionality reduction, clustering, and batch correction in multi-omics integration, leveraging both graph-based knowledge and adversarial alignment for effective data fusion across unpaired multi-omics datasets.

**MultiVI** [148] already described, also employs a classifier network to classify samples of the shared latent space into batches/modalities. The correspondent cross-entropy loss is adversarially trained to minimise batch effects, penalising the model if samples from different modalities are overly separated in the latent space.

**scCross** [150] is a recent model designed for cross-modality translation, multi-omics data simulation, and to perform *in silico* perturbations. For cross-modality translation, measurements from one omics modality are mapped into a shared latent space using the modality-specific encoder, which is then passed through the decoder of another omics modality. scCross trains modality-specific VAEs to extract low-dimensional cell embeddings that are converted to a shared latent space, using an FFNN aligner. A global discriminator operates on the shared latent space to identify the omics origin of the cells, ensuring proper alignment and integration across modalities. To refine cross-modality generation and ensure accurate data reconstruction, modality-specific discriminators are employed. These discriminators try to distinguish between the original and reconstructed data for each omics modality, maintaining data integrity and consistency during reconstruction. scCross is one of the most recent methods for omics cross-modality translation. However, other notable models for this task include **scMM** [132], **PolarBear** [162] and **Portal** [144].

## Cycle-consistency training

Other regularisation terms can be included in the loss function of a VAE, such as cyclical consistency terms for additional intra-modal and cross-modal consistency inspired in cycleGAN [163]. For intra-modal consistency, the latent space of one modality is decoded and then re-encoded with the modality-specific encoder. These re-encoded embeddings are compared to the original ones to align the two latent spaces. For cross-modal consistency, low-dimensional embeddings from one modality are decoded and subsequently re-encoded with the decoder and encoder of another modality *(Figure 3d)*. By aligning these cross-modal embeddings with the original ones, the model can learn to produce cross-modal translations that are consistent with the original sample when re-embedded in the latent space.

**scMVP** [142] is a MVAE for di-omics vertical integration. It employs modality-specific encoders and decoders and a GMM prior to derive the shared latent space. This model uses multi-head self-attention transformer modules for scATAC-seq data to highlight the most informative features [131]. Simpler attention blocks are used for scRNA-seq data to dynamically weight features, emphasising their importance during training. scMVP integrates single-modal encoders with the joint latent space to ensure clustering consistency by minimising the KL-divergence between the joint latent embeddings and the modality-specific re-embeddings from the decoder output:

$$L_{scMVP-consistency} = D_{KL}(q(z|x_1, x_2)||q(z|\widehat{x_1})) + D_{KL}(q(z|x_1, x_2)\big|\big|q(z|\widehat{x_2}))$$  (15)

Where $x_1, x_2$ are two omics modalities sample vectors, $\widehat{x_1}$ and $\widehat{x_2}$ are the correspondent reconstructed omics vectors. This cyclical intra-modal consistency loss ensures robust integration, the alignment of omics reconstructions and is used to impute missing data.

The application of cycle-consistency regularisers extends beyond VAE models and is utilized in various architectures, including autoencoders and heterogeneous frameworks. For example, **con-AAE** [164] is an autoencoder designed for diagonal integration tasks with a cycle-consistency loss term. It uses two autoencoders to map each modality into its own low-dimensional embedding, working to unify and align these embeddings through an adversarial loss to distinguish samples from the latent spaces, and a latent cross-modal cycle-consistency loss *(Figure 3d)*. The cross-modal cycled embedding generated is compared to the original latent representation to ensure consistency through an additional cycle-consistency term in the autoencoder loss:

$$L_{conAAE-consistency} = \mathbb{E}_{x_1}[d(E_1(x_1), E_2 D_2 E_1(x_1))] + \mathbb{E}_{x_2}[d(E_2(x_2), E_1 D_1 E_2(x_2))] \qquad (16)$$

Where $E_1$, $D_1$ and $E_2$, $D_2$ are the encoders and decoders for omics modalities $x_1$ and $x_2$, respectively, and $d$ stands for indicated distance in the embedding space. This approach enforces alignment between the latent spaces of both modalities, encouraging cross-modal similarity.

On the other hand, **sciCAN** [145] is a heterogeneous model for the integration of two omics modalities that combines representation learning using an encoder with modality alignment through a cycle-GAN. The encoder projects both modalities into a shared low-dimensional latent space, using noise contrastive estimation as the loss function to retain the intrinsic structure of the data. The GAN component includes two discriminators to ensure modality alignment. One discriminator distinguishes between latent spaces of the two modalities, forcing the encoder to minimise modality-specific differences. The second discriminator distinguishes between real and generated data from the cross-modal generator. sciCAN also employs a cycle-consistency loss, as described in the previous sections, to align the embeddings of the encoded generated data with the original ones.

**Portal** [144] is another example that integrates two omics modalities using two encoders, generators and discriminators. The encoders learn latent embeddings for each omics modality and the cross-modal generators generate synthetic omics data. The discriminators distinguish between original and generated data and also between omics unique or shared cell types. Portal incorporates additional consistency regularisers: an autoencoder loss for intra-modality reconstructions, a latent alignment loss for consistency between the cross-modal embedding and the original one, and a cosine similarity loss for correspondence between the original and cross-modal reconstructed samples.

Adversarial and cycle consistency training strategies have proven effective in aligning and integrating multi-omics data by encouraging consistency across modalities, improving reconstruction accuracy, and enabling the translation between different omics modalities.

## Contrastive learning

Contrastive learning offers an alternative and complementary approach to extract meaningful representations in unsupervised learning methods. The main idea is to ensure that similar data points (positive pairs) are represented closer together in the latent space, while dissimilar data points (negative pairs) are pushed further apart [165]. In the context of multi-omics integration and alignment, contrastive learning can be applied to reduce distances between instances from the same cluster, pushing them towards each other, and to increase distances between instances from different clusters which are forced away [166]. By enforcing these relationships during training, the model effectively learns to align similar sample types across modalities, improving accuracy and robustness in downstream tasks *(Figure 3e)*. To distinguish the positive pair from the negatives, several metrics and functions can be used [23,164,166,167].

**MOSA** [23] is a conditional MVAE model that adopts the concatenation strategy to build the joint latent space. This model added a self-supervised contrastive loss to the standard VAE loss, defined as:

$$L_{cosine-contrastive} = [m_{pos} - s_p]_+ + [s_n - m_{neg}]_+ \tag{17}$$

where $s_p$ and $s_n$ represents the cosine similarity between positive pairs and negative pairs defined by whether two samples have the same tissue type, and $m_{pos}$ and $m_{neg}$ are positive and negative margins, also tuned during model training. This model is capable of integrating seven distinct bulk omics datasets, including drug response and CRISPR-Cas9 screen data. Additionally, key biological features such as cancer driver mutations are concatenated as conditionals to each omics layer prior to encoding and to the latent neurons before decoding, enhancing model's reconstruction and biological relevance. By incorporating SHAP for model interpretation, it identifies key multi-omics features critical for cell clustering and for uncovering biomarkers associated with drug sensitivity and gene dependencies.

**con-AAE** [164] previously described also incorporates a contrastive loss by taking advantage of the ground-truth cell type labels. The additional regulariser aims to minimise:

$$L_{conAAE-contrastive} =$$
$$\mathbb{E}_{x_1}[d(E_1(x_1), z^p) - d(E_1(x_1), z^n) + \alpha] + \mathbb{E}_{x_2}[d(E_2(x_2), z^p) - d(E_2(x_2), z^n) + \alpha] \tag{18}$$

Where $d$ is an indicated distance in the embedding space, $z^p$ is the hard positive vector in the embedding space which is the furthest vector from the defined anchor vector within the same cluster, $z^n$ is the hard negative, defined as the closest vector from a different cluster, and $\alpha$ is the margin hyperparameter tuned when training the model.

By aligning similar samples and separating dissimilar ones, contrastive learning refines the representation of shared structures across omics modalities. However, to further unravel the independent and interpretable factors driving biological variation, disentanglement learning emerges as a complementary approach.

## Disentanglement learning

Disentanglement representation learning (DRL) [168] is a machine learning strategy designed to learn latent representations that can separate the underlying independent and informative factors of variation in the data. In multi-omics integration, DRL is particularly powerful for disentangling the complex and heterogeneous molecular processes, enabling a clearer interpretation of biological systems. VAEs are particularly effective for DRL, as their flexible architecture allows them to incorporate regularisers to encourage disentanglement. This approach enhances model interpretability, robustness, and generalizability, making it valuable for multi-omics analysis and batch effect correction. Several disentanglement regularisers for VAE and GAN approaches are reviewed in [168].

**scDisInFact** [149] is a VAE framework designed to learn latent factors that disentangle condition from batch effects in scRNA-seq data, enabling it to simultaneously remove batch effects and identify condition-associated key genes. The goal is to disentangle shared-bio factors ($z_s$) and unshared-bio factors ($z_u$) through a combination of loss terms and encoder-specific strategies. An additional MMD term plays a pivotal role in disentanglement by ensuring $z_s$ is independent of condition and batch labels and is expressed as:

$$L_{MMD-disentanglement}(z_s) = \sum_{c=1}^{C} \sum_{i \in B_c} MMD(z_s^{ref} || z_s^i) \tag{19}$$

where $B_c$ is the set of batches under condition label $c$, and $C$ the total number of conditions. $z_s^{ref}$ is the latent representation of a reference batch and condition and $z_s^i$ represents the remaining ones. A MMD term is also applied to $z_u$ to enforce independence from batch effects while preserving condition-specific variability:

$$L_{MMD-disentanglement}(z_u) = \sum_{c=1}^{C} \sum_{i \in B_c} MMD(z_u^{ref(c)} || z_u^i) \qquad (20)$$

where $z_u^{ref(c)}$ is the latent representation of a reference batch under condition label $c$. For datasets with multiple condition types, this is expanded to include all unique combinations of conditions and batches. The disentanglement is reinforced by the ELBO loss, which combines reconstruction and KL divergence terms, MMD terms, group lasso loss for feature selection and cross-entropy loss for condition prediction using $z_u$.

**MIDAS** [51] is a MVAE for mosaic integration of single-cell multimodal data that uses self-supervised learning and information-theoretic latent disentanglement [169] to achieve dimensionality reduction, imputation, and batch correction. The total VAE loss comprises three components: the ELBO, a disentanglement regulariser and an alignment term. The disentanglement term is based on the Information Bottleneck loss and aims to disentangle the shared latent space into biological states $c$ and technical noise $u$. MIDAS also highlights the need for approaches that can effectively handle missing modalities in mosaic multi-omics data.

Missing modalities in mosaic multi-omics data

Mosaic integration methods address the challenges of combining multi-omics datasets with incomplete and overlapping modalities across samples *(Figure 1)*. These methods aim to overcome limitations in scalability, cost, and modality coverage inherent in existing sequencing technologies. However, these are the most challenging models due to some obstacles that include reconciling modality heterogeneity, managing technical variation across batches, and enabling robust modality imputation and batch correction for downstream analyses. Recent approaches, some of them already described in this review, provide promising solutions for this problem based on matrix factorization [74,170], VAEs [49–51,146,171], or other approaches such as **StabMap** [172].

For example, **Multigrate** [49] previously described takes the PoE approach and deals with the missing data by setting the posterior of the missing modality to 1. Therefore, the joint latent distribution can still be determined from the modalities that are present while ignoring missing modalities. This enables the generation of a joint embedding regardless of a sample's modality-missing pattern and the reconstruction of all modalities, even if some were missing in the input.

**scVAEIT** [146] models missing features and modalities in multi-omics datasets through conditional variational inference. An actual mask identifies the true missing data pattern. During training, random masks $M$ are generated and encoded to encourage the model to predict the missing features based on the remaining observed data $X_{M^c}$. An encoder learns the parameters for the posterior distributions of the latent space based on $M$ and $X_{M^c}$. The decoder predicts the posterior mean of the original dataset $X$ based on the latent space constructed, on the mask $M$ and sample covariates. This process allows the imputation of the unobserved values and the denoising of the observed features.

**MIDAS** [51] also takes the PoE approach and the missing features and modalities are addressed using padding and masking techniques. For each cell, MIDAS pads the missing features with zeros, ensuring that each feature vector has the same fixed size. The learned joint disentangled latent variables are passed to each modality decoder, and a masking function is used to remove the padded missing features from a generated padded feature mean vector. This returns the imputed values for the missing features and can be used for downstream analysis.

The flexibility of VAEs underscores their key role in advancing deep learning-based multi-omics integration, addressing challenges like handling missing modalities and enabling interpretable latent spaces. VAEs have shown highly adaptability through innovative strategies such as adversarial training, cycle consistency, contrastive learning, and supervision for specific tasks. While these models provide robust frameworks for extracting meaningful biological insights, further refinement of their architectures is essential to enhance scalability, broaden applicability, and maximize their potential impact on precision medicine and synthetic biology.

## Promising perspectives

Classical statistical and machine learning approaches have proven effective for integrating multi-omics data, as highlighted throughout this review. As the field progresses, several emerging themes are becoming increasingly prominent, suggesting promising directions for future developments in multi-omics integration. These include the integration of other data modalities alongside molecular omics and the development of foundation models. These innovations offer the potential to extend the scope and accuracy of multi-omics integration, facilitating the construction of more comprehensive and robust models of biological systems.

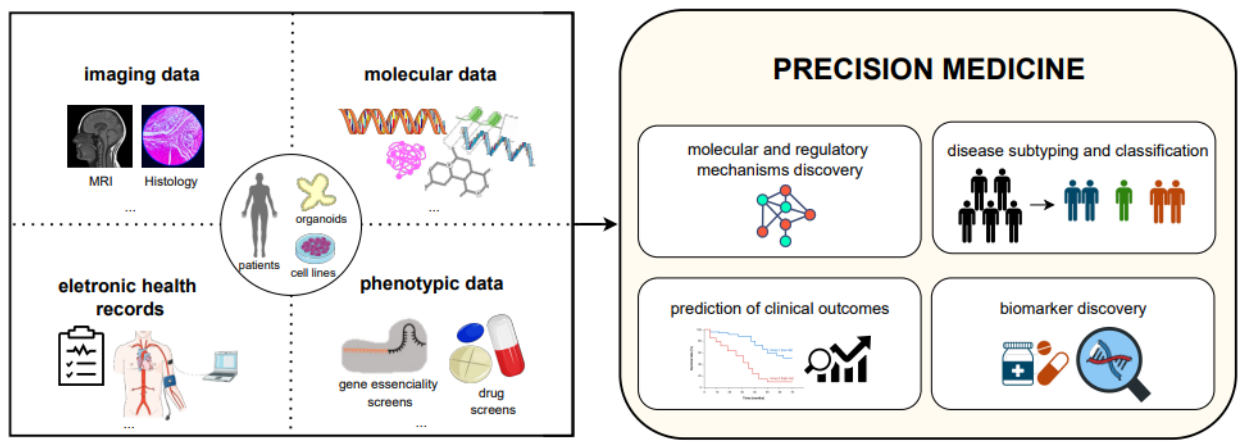

**Figure 4.** *Integration of multiple data modalities - molecular, phenotypic, imaging and electronic health records datasets (left) - to uncover molecular and regulatory mechanisms, enable disease subtyping and classification, enhance clinical outcome predictions for diagnosis and prognosis, and identify biomarkers of therapeutic response (right).*

## Beyond omics: multi-modal integration

The rapid advancement of biomedical technologies and the increasing diversity of data modalities present unprecedented opportunities for precision medicine and synthetic biology. Beyond molecular omics data, modern approaches increasingly integrate phenotypic datasets, imaging modalities (e.g., histopathology slides, MRI, PET), electronic health records (EHRs), and bio-signals from wearable devices. Additionally, experiments generating these data modalities are conducted in various biological systems, including cell lines, organoids, and patient-derived samples, offering diverse experimental contexts. Phenotypic datasets, such as CRISPR-Cas9 and drug-response screens, are invaluable for identifying genetic dependencies and therapeutic vulnerabilities, linking molecular profiles to functional biological effects [7,13,23,173–175]. Imaging data provides essential spatial and morphological context, particularly relevant in fields such as oncology and neurodegenerative diseases. EHRs, including clinical history or laboratory results, link molecular and phenotypic data to real-world patient outcomes, while wearable devices continuously monitor health metrics in real-time.

The integration of diverse data modalities provides an even more holistic view of biological processes and disease mechanisms. To fully harness this potential, several models and architectures previously described, such as VAEs, along with other advanced artificial intelligence (AI) methodologies, are essential for the effective integration of these heterogeneous datasets. AI-based approaches have already demonstrated success in various applications, as highlighted in [176–178]. By leveraging these computational methods, it will be possible to integrate and model all available data more effectively, paving the way for the identification of multimodal biomarkers and advancing precision medicine through more personalised and data-driven patient care.

## Transformers and Foundation Models

A rapidly growing area in AI is the development of foundation models, originally designed for natural language processing [179]. These models, typically based on the self-attention transformer architecture, are pre-trained on large and diverse datasets, enabling them to generalize across various domains. By leveraging transfer learning, a pre-trained model allows for fine-tuning to a specific domain or task with much less data than would be needed to train a model from scratch. Notable examples include BERT [180] and GPT-4 [181] that have achieved remarkable success in fields like computer vision, speech recognition and natural language generation, consistently outperforming task-specific models.

Foundation models are now being expanded to biological datasets, leveraging their ability to handle heterogeneous datasets and to generalise across multiple tasks. Furthermore, the attention mechanisms inherent in transformer architectures enhance interpretability by identifying critical features or relationships within biological datasets. Recent foundation models [182–186] exemplify their use in biology, demonstrating versatility in multi-batch and multi-omics integration, perturbation response prediction, tissue drug response prediction, cell type annotation or gene regulatory network inference tasks. Other transformers-based models for single-cell omics are reviewed elsewhere [187]. Recent research has also explored the potential of these models to integrate multi-omics data with complex biological networks [188,189] or with specific pathway information, such as the transformer-based DeePathNet [190]. Albeit, applying these models to omics analyses presents challenges [191,192]. One notable limitation is the lack of inherent sequential structure in omics data, which requires strategies such as gene order ranking to address this issue. Additionally, concerns have been raised regarding the application of foundation models compared to classical machine learning approaches, highlighting the need for further refinement of these models and careful identification of their most relevant applications.

## Conclusions

The integration of multi-omics data has revolutionised the study of complex biological systems, providing comprehensive insights into molecular mechanisms and advancing precision medicine. This review categorizes multi-omics integration methods into their underlying approach, providing a comprehensive technical perspective on the models developed.

Classical approaches remain highly effective for datasets with limited sample sizes, offering robust and interpretable tools for tasks like dimensionality reduction, clustering, and pathway analysis. On the other hand, deep learning models have significantly advanced the field by enabling the integration of high-dimensional, incomplete, complex, and heterogeneous data, although they often require large sample sizes for effective training. Among deep learning approaches, VAEs stand out for their flexibility and generative capabilities. For instance, VAEs can condition the latent space on discrete genomic data to integrate and represent the heterogeneity of omics data. Furthermore, regularisation terms can be incorporated into the common loss function to address unique challenges such as adversarial terms for batch correction and disentanglement or contrastive terms to enhance interpretability and extract biologically meaningful representations. These capabilities make VAEs particularly powerful tools in multi-omics integration.

Foundation models represent a groundbreaking development to address the challenges of sample size in deep learning. As computational power increases and datasets grow, integrating foundation models, leveraging transfer learning, and incorporating prior biological knowledge will be crucial for advancing multi-omics research, accelerating the understanding of complex diseases, and enabling the development of more personalized therapeutic strategies.

# Acknowledgements

ARB, SV and EG work is partially supported by national funds through FCT, Fundação para a Ciência e a Tecnologia, under projects UIDB/50021/2020 (DOI:10.54499/UIDB/50021/2020) and LAETA (DOI:10.54499/UIDB/50022/2020). ARB is also funded by the Portuguese national agency FCT, through the research grant UI/BD/154599/2022. Rebecca Poulos is supported by a Sydney Cancer Partners Translational Partners Fellowship with funding from a Cancer Institute NSW Capacity Building Grant (Grant ID. 2021/CBG0002).

## Data availability

No new data were generated or analysed in support of this research.

scale atlas-level single-cell datasets. Nat. Comput. Sci. 2022; 2:317–330